\documentclass{elsart}
\usepackage{graphicx}

\begin{document}
\begin{frontmatter}
\title{Guiding and reflecting light by boundary material}
\author[oma]{A.\ Huttunen\corauthref{cor1}},
\author[karri]{K. Varis}
\author[vtt]{K.\ Kataja},
\author[vtt]{J.\ Aikio},
\author[paivi]{P.\ T\"orm\"a}
\address[oma]{Department of Electrical and Communications Engineering, 
Laboratory of Computational Engineering, Helsinki University of Technology, 
FIN-02015 HUT, Finland}
\address[karri]{Department of Electrical and Communications Engineering, 
Optoelectronics Laboratory, Helsinki University of Technology, 
FIN-02015 HUT, Finland}
\address[vtt]{VTT Electronics, FIN-90571 Oulu, Finland}
\address[paivi]{Department of Physics, Nanoscience Center, 
FIN-40014 University of Jyv\"askyl\"a, Finland}
\corauth[cor1]{Corresponding author.
Tel: +358 9 451 5730;
Fax: +358 9 451 4830;
Email: anu.huttunen@hut.fi}

\begin{abstract}
We study effects of finite height and surrounding material on
photonic crystal slabs of one- and two-dimensional photonic crystals 
with a pseudo-spectral method and finite difference time domain 
simulation methods. The band gap is shown to be strongly modified by 
the boundary material. As an application we suggest reflection and 
guiding of light by patterning the material on top/below the slab.

\end{abstract}


\end{frontmatter}

Photonic crystals are periodic dielectric structures that exhibit 
a photonic band gap, which can be utilized to control and confine 
light~\cite{joannopoulos_book}. Photonic crystal slabs are planar 
structures, i.e., they are periodic in a plane, but have finite 
thicknesses in the third dimension. The slab acts as a planar waveguide, 
if the average refractive index of the slab is higher than that of 
the surroundings. Thus light can be controlled three-dimensionally, 
when the slab is patterned to form a two-dimensional photonic 
crystal~\cite{johnson99,chow00}. Two-dimensional photonic crystal slabs 
are considerably easier to fabricate than three-dimensional photonic crystals. 
Analogously, light can be 
controlled two-dimensionally with one-dimensional photonic crystal 
slabs~\cite{fan95,atkin96,patrini02,koshino03}.
The characteristics of photonic crystal slabs are 
notably different from the corresponding infinite photonic crystals and
thus miniaturized photonic crystal components for, i.e., integrated optics 
cannot be designed based on the features of infinite structures.

Waveguides are basic components of an integrated optical circuit.
Waveguides in two-dimensional photonic crystal slabs are formed by 
introducing line defects to the periodic structure~\cite{kuchinsky00,loncar00,lin00,johnson00,loncar00_2,smith00,chow01,baba01,notomi01,chutinan02}. 
Light with a frequency in the band gap of the photonic crystal cannot 
propagate in the periodic structure and thus follows the line defect.
We are proposing a new way to guide light using the influence that 
the material surrounding the photonic crystal slab has on the band gap.
The photonic crystal slab has a band gap for different frequencies 
depending on the material that is on top/below the slab. 

We consider photonic crystal slabs of 
heights $h$ of the order of the period $P$. 
We calculate the band structures of two-dimensional photonic crystal slabs 
using a pseudo-spectral method~\cite{karri}.
Here, we show calculations for a specific geometry
to illustrate the effects of different types of boundary 
materials above and below a photonic crystal slab. 
With boundary material we refer to the half-spaces above and below the slab.
We show that varying the boundary material results in large changes in the 
band structures. We suggest
to use this effect for guiding of light by patterning the boundary
material --- an alternative way for light guiding along crystal defects.
The idea is the following (also explained in Fig.\ \ref{2dwg}): 
planar two-dimensional photonic crystal would be fabricated with no defects
and the waveguide would be realized by covering a channel
(or leaving a channel uncovered) in otherwise uncovered (covered) slab. 
The difference of the band structures in covered/uncovered areas
confines light in the waveguide. We demonstrate the idea with 
three-dimensional finite difference time domain (FDTD) simulations.
For one-dimensional photonic crystal slabs the same effect can be used
for reflecting light traveling along a one-dimensional photonic crystal 
slab by changing the boundary material abruptly. We demonstrate this
with a two-dimensional FDTD method.
The FDTD methods are described in Ref.\ \cite{taflove}.

\begin{figure}
\centering
\includegraphics[scale=0.6]{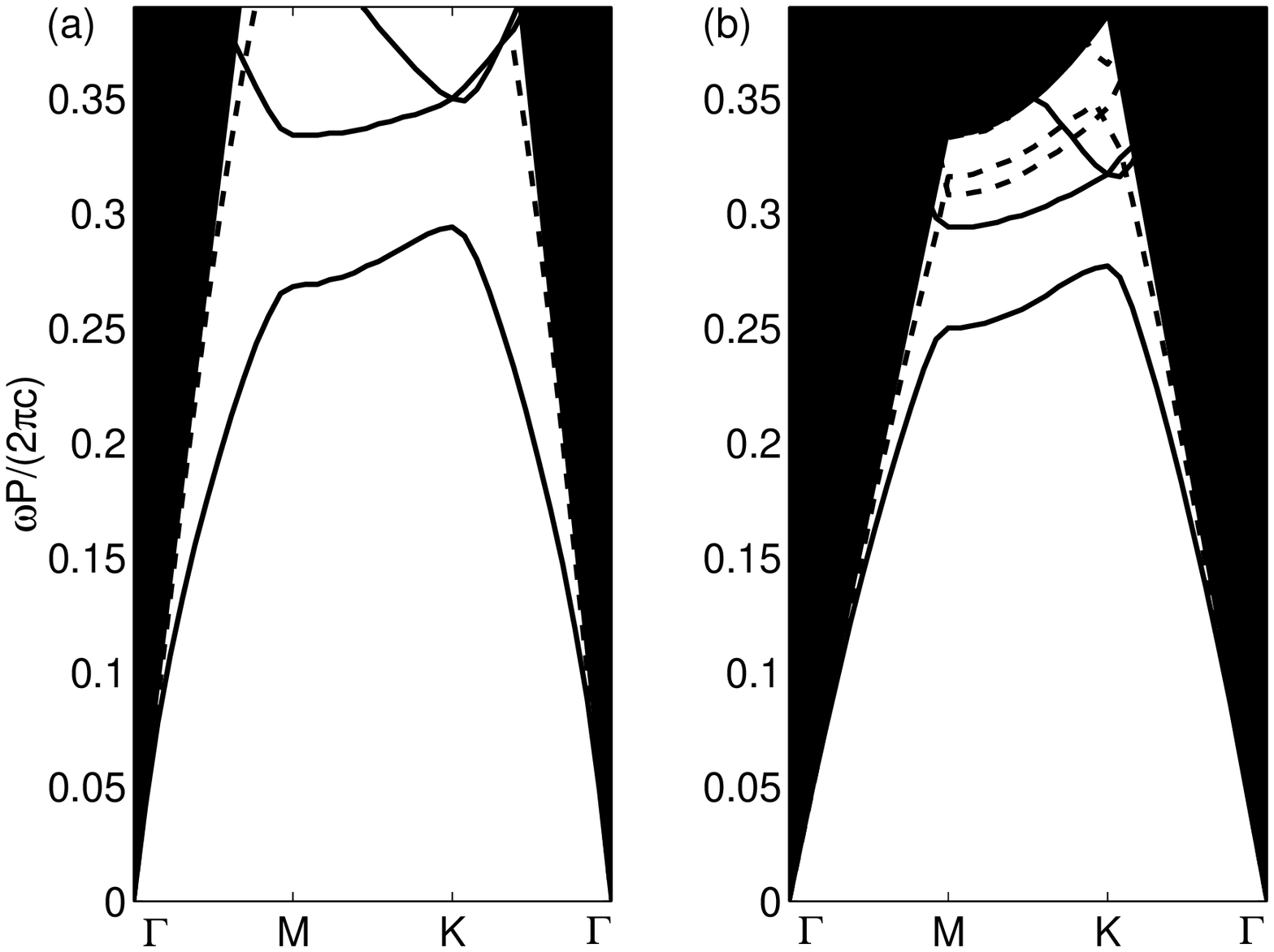}
\caption{Band structures of a two-dimensional photonic crystal slab with 
triangular lattice of air holes. 
The dielectric constant of the slab is $\varepsilon=12$. 
The boundary material on top/below the slab is air $\varepsilon_b=1$ in (a) 
and dielectric $\varepsilon_b=3$ in (b).
The lattice parameters are radii of the holes $r/P=0.24$ and the height 
of the slab $h/P=0.3$ with respect to the period $P$.
Solid curves are even modes, dashed curves are odd modes and 
black areas are above the light lines 
for the boundary material in question.}
\label{2dbs}
\end{figure}

\begin{figure}
\centering
\includegraphics[scale=0.6]{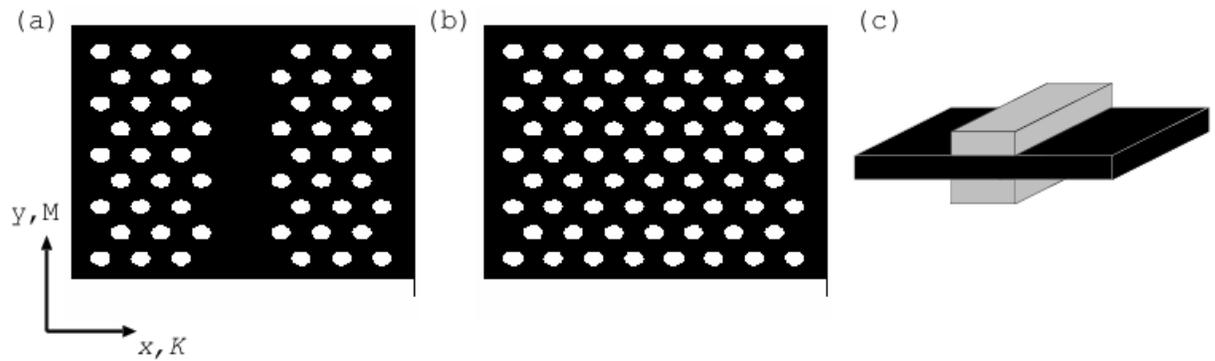}
\caption{Geometries for the 3D FDTD simulation of wave guiding 
in two-dimensional photonic crystal slabs. 
(a) Conventional photonic crystal waveguide: guiding along a line 
defect of missing air holes.
(b) Slab of triangular lattice.
(c) Guiding by boundary material: the slab with the geometry shown in (b) 
is drawn in black. Stripes of material with dielectric 
constant $\varepsilon_{wg}=3$ are placed on top and below the slab 
(drawn in gray). 
The parameters of the lattice are as in Fig.~\ref{2dbs}. 
}
\label{2dwg}
\end{figure}

To demonstrate that boundary material affects the band structure, we
calculated band structures of two-dimensional photonic crystal slabs 
using a pseudo-spectral method~\cite{karri}. 
The band structures for a triangular lattice photonic crystal slab with 
boundary materials air and a dielectric, with a dielectric constant 
$\varepsilon_b=3$, are shown in Fig.\ \ref{2dbs}.
It can be seen that when the boundary material is air, the frequencies
around $\omega P/(2\pi c)=0.3$ fall into the band gap for even modes 
in the M-direction [Fig.\ \ref{2dbs} (a)].
However, when the boundary material has $\varepsilon_b=3$, there
is an even mode at $\omega P/(2\pi c)=0.2937$ at the M-point 
[Fig.\ \ref{2dbs} (b)].
Thus a waveguide can be made for this frequency in the M-direction by 
covering the photonic crystal slab in the waveguide region and leaving it 
otherwise uncovered. 
Light will follow the covered region where an allowed mode exists.

\begin{figure}
\centering
\includegraphics[scale=0.4]{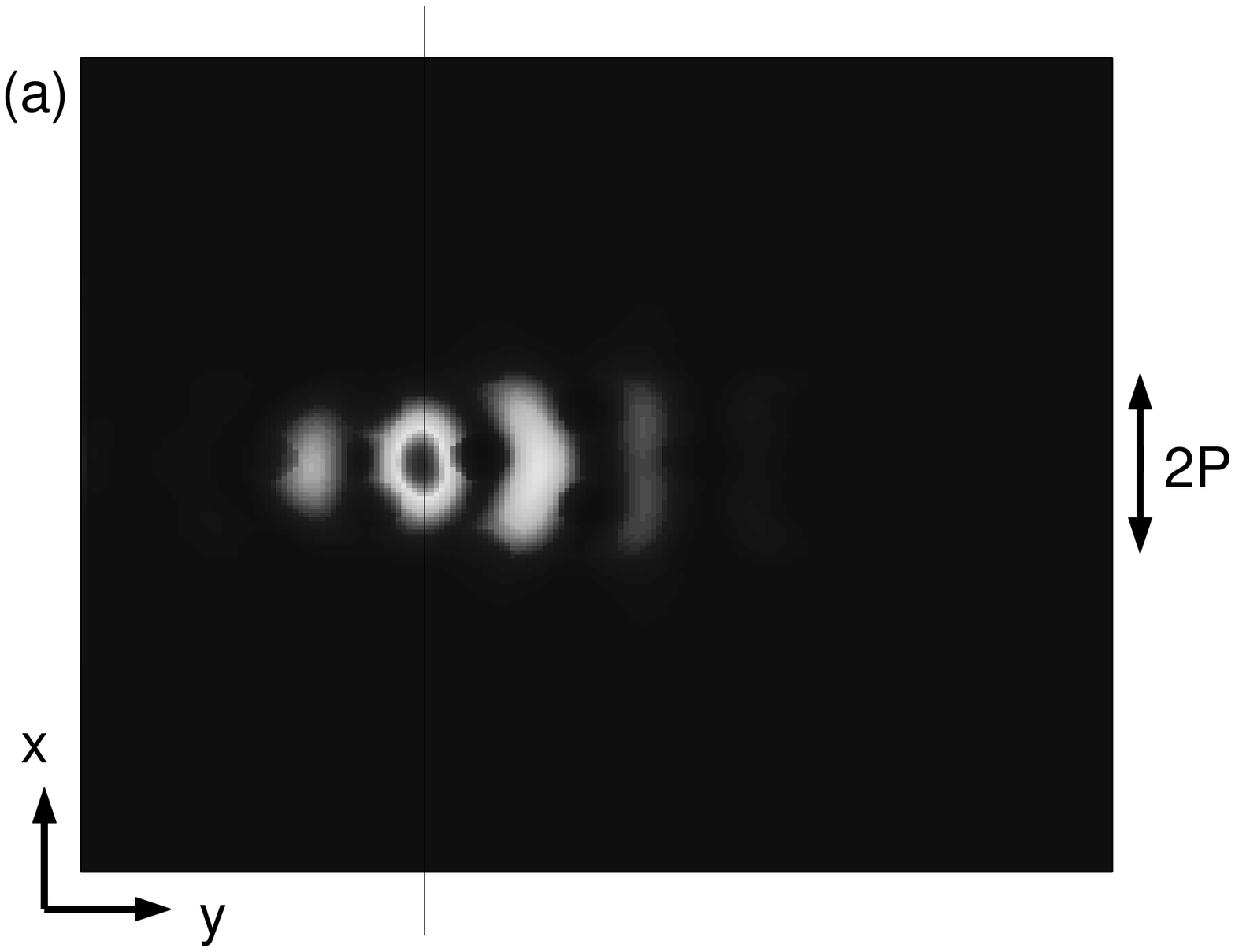}
\includegraphics[scale=0.4]{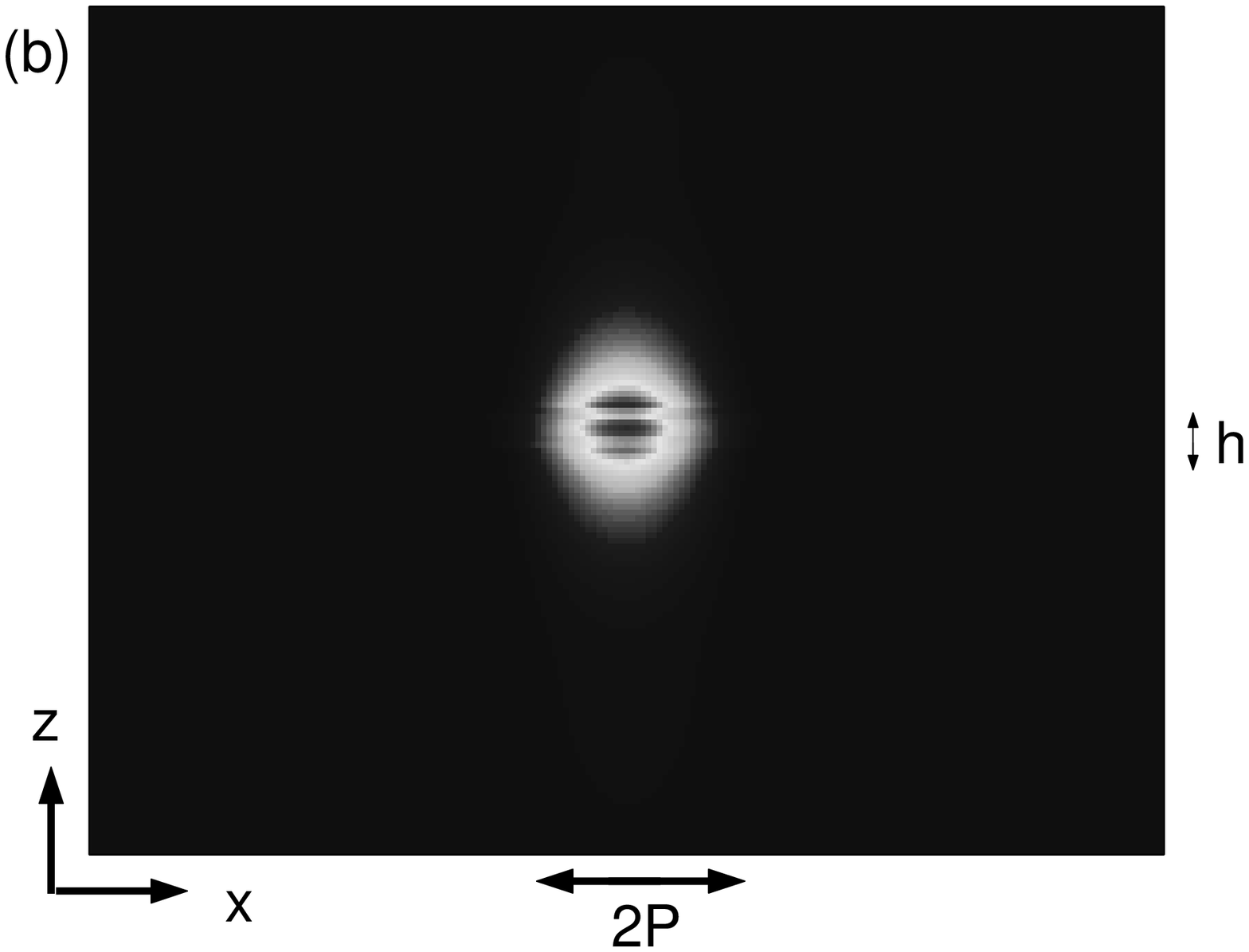}
\caption{The energy density of the Gaussian pulse propagating in 
the waveguide with the geometry shown in Fig.\ 2 (b)-(c). 
In (a), the pulse is propagating in the $y$-direction. The energy density 
is integrated over the waveguide height in the $z$-direction.
The cross section of the energy density at the plane denoted 
by a line is presented in (b). 
The waveguide width (2P) is denoted in both figures. }
\label{energydensity}
\end{figure}

We performed three-dimensional FDTD calculations to study this
type of waveguiding using the boundary material. 
The structures are shown in Fig.\ \ref{2dwg}. 
We simulated the propagation of a Gaussian pulse with wavelength
$\lambda =1500$ nm. The period of the photonic crystal lattice was
$P=450$ nm, so that $P/\lambda =0.3$. 
The spectral full width at half 
maximum (FWHM) of the pulse was $\Delta \omega P /(2\pi c) =0.1$, so 
that the main frequencies of the pulse fit into the band gap of 
Fig.~\ref{2dbs} (a). 
First we calculated the transmission of 
the pulse along a defect waveguide made by removing 
holes from the photonic crystal slab [Fig.\ \ref{2dwg} (a)].
Then we calculated the transmission for a photonic crystal slab with no 
defects [Fig.\ \ref{2dwg} (b)] which was covered by strips of dielectric 
with $\varepsilon_{wg}=3$ [Fig.\ \ref{2dwg} (c)].
The width of the strips was $2P$.
The  transmission of a Gaussian pulse for the guiding by boundary material 
[Figs.\ \ref{2dwg} (b) and \ref{2dwg} (c)] was 40\% larger than in the 
traditional defect waveguide case [Fig.\ \ref{2dwg} (a)].
The energy densities of a Gaussian pulse propagating in the waveguide
is presented in Fig.~\ref{energydensity}.

\begin{figure}
\centering
\includegraphics[scale=0.6]{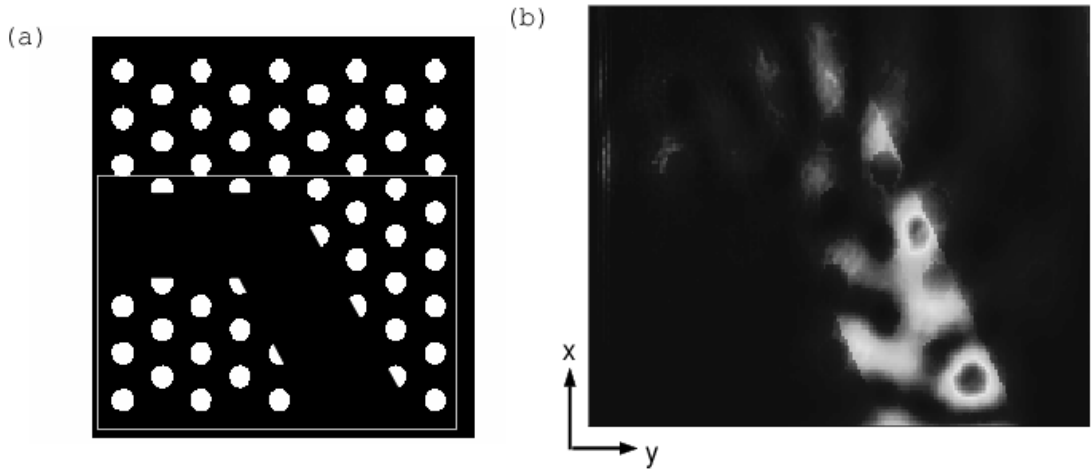}
\caption{(a) Geometry of a 60$^{\circ}$ turn in the hexagonal lattice 
photonic crystal. The black waveguide areas are realized by placing 
waveguide strips on top/below the photonic crystal slab with no embedded 
defects. The geometry parameters are the same as in Figs.~\ref{2dbs} 
and \ref{2dwg}. (b) The energy density of the Gaussian pulse after 
propagating through the turn. The energy density is integrated over 
the waveguide height in $z$-direction and the shown area in $xy$-plane 
is marked with a white box in (a). The spectral full width at half 
maximum (FWHM) of the pulse was $\Delta \omega P /(2\pi c) =0.056$.}
\label{turn}
\end{figure}

We studied also 60$^\circ$ and 90$^\circ$ turns in the hexagonal 
lattice. For the 60$^\circ$ turn (see Fig. \ref{turn}), it was found 
that 51-55\% of the light follows the waveguide turn. (Throughput of 51\% 
is reached for FWHM $\Delta \omega P /(2\pi c) =0.1$ and 55\% for 
FWHM $\Delta \omega P /(2\pi c) =0.056$. The width of the band gap is 
$\Delta \omega P /(2\pi c) =0.066$.) In contrast, only a 
small fraction (about 10\%) of the light followed the 90$^\circ$ turn 
because there is no solution in the K-direction for the used wavelength 
in the covered photonic crystal ($P/\lambda=0.3$). We repeated the 
simulations for a system where the photonic crystal was replaced by a 
material with a refractive index corresponding to the average index of 
the photonic crystal material (n = 2.9493): the amount of guided light was 
similar. Therefore, for this geometry and material, the effect of the 
photonic crystal reduces to an average refractive index material in a 
conventional ridge waveguide configuration. Finally, we repeated the 
simulations of the 60$^\circ$ and 90$^\circ$ turns for photonic crystal 
defect waveguides, and the guiding efficiency was considerably smaller 
(of the order of few percents). Note that the geometries 
and the coupling into the waveguides were not optimized, thus the 
results could probably be improved by optimizing the parameters. 
Furthermore, the comparison to ridge and defect waveguides refers to 
this specific example and should not be generalized without further 
study. However, the results demonstrate the in-principle usability of 
the idea.

\begin{figure}
\centering
\includegraphics[scale=0.5]{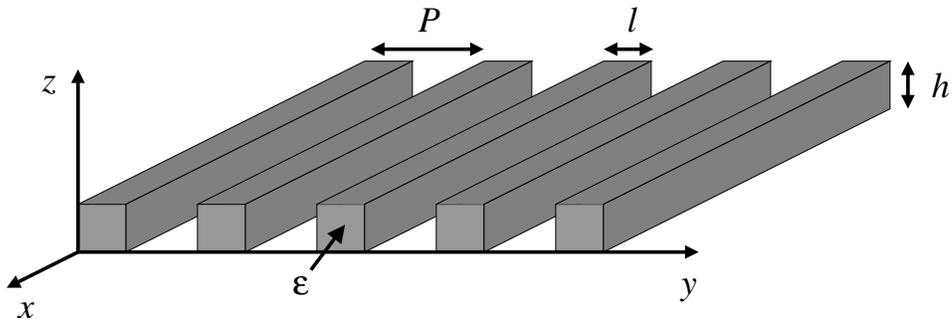}
\caption{Geometry of a one-dimensional photonic crystal slab. 
The rods, which are infinitely long in the $x$-direction and have a finite
thickness $h$ in the $z$-direction, are periodically replicated in
the $y$-direction. The parameters are thickness of one layer $l/P=0.2$ 
and height of the slab $h/P=0.5$, relative to the period $P$, and dielectric 
constants of the materials. Here, the other material in the periodic stack 
has dielectric constant $\varepsilon=13$ and the other material is air.
The dielectric constants of the boundary material $\varepsilon_b$ 
above ($z>h$) and below ($z<0$) the slab are varied in the simulations.}
\label{crystal}
\end{figure}

\begin{figure}
\centering
\includegraphics[scale=0.6]{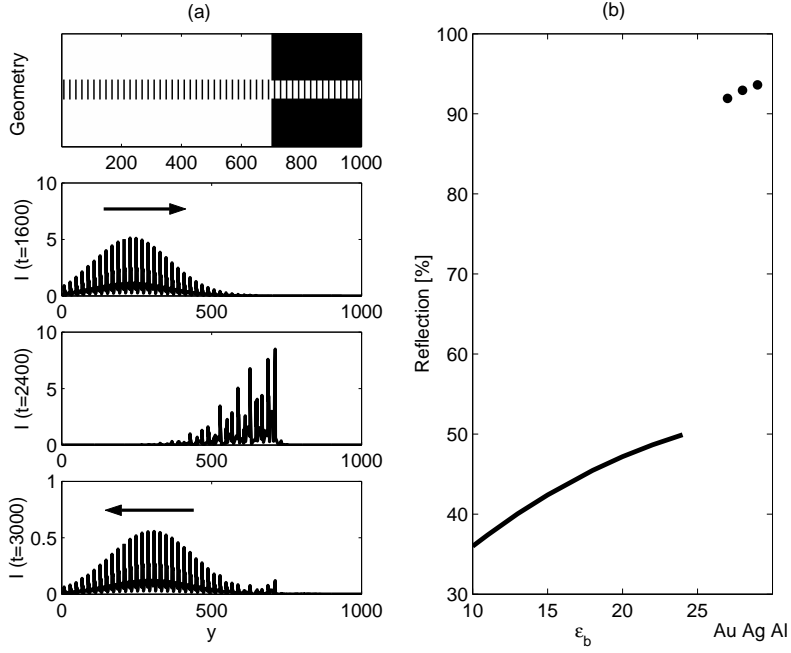}
\caption{(a) Energy density profiles of a Gaussian pulse in 
a one-dimensional photonic crystal slab at different times. 
The profile is taken in the middle of the photonic crystal. 
Photonic crystal geometry is shown in the upmost part of the figure. 
The black areas denote GaAs and white ones air. The pulse has a scaled 
frequency $\omega P/(2\pi c)=0.3$. Light with this frequency can 
propagate in the photonic crystal when the boundary material is air, 
but falls into the band gap when the boundary material is GaAs. 
This can be seen from the intensity profile as the pulse is reflected.
(b) The fraction of the energy density of a Gaussian pulse that is 
reflected inside the slab from the point where the boundary material above 
and below the photonic crystal slab changes
from air to a material with dielectric constant $\varepsilon_b$.
The rest of the energy of the pulse is diffracted into the boundary material.
Solid line indicates dielectric boundary materials and dots indicate metals:
gold, silver, and aluminum for which the refractive indices are
$n_{Au}=0.053+7.249i$, $n_{Ag}=0.078+7.249i$, and $n_{Al}=0.384+11.88i$, 
respectively.}
\label{fdtd}
\end{figure}

The one-dimensional photonic crystal slab is a periodic stack of dielectric 
rectangular rods which have different dielectric constants.
The considered geometry is shown schematically in Fig.\ \ref{crystal}.
Parameters needed to define the structure are explained in the figure 
caption. Light traverses the structure ($y$-direction in Fig.\ \ref{crystal}),
it reflects from each layer and interferes resulting in a band structure.
We consider a polarization with field components $H_x$, $E_y$, and $E_z$.
As discussed above, the band gap of a photonic crystal slab depends on 
the dielectric constant of the boundary material. In structures with 
periodicity in one dimension only, it can be utilized for reflecting 
light by changing the boundary material abruptly. For the considered 
structure, light with frequency around $\omega P/(2\pi c)=0.3$ has a 
solution in the photonic crystal when the slab is in air, but falls 
into the band gap when the slab is sandwiched between dielectric 
$\varepsilon_b=13$. We can see from the two-dimensional 
FDTD simulations that light reflects at the point where the boundary material 
changes [see Fig.\ \ref{fdtd}(a)]. Part of the light is diffracted to the 
boundary material that has a high dielectric constant. To estimate the 
efficiency of the reflection, we calculated the energy density of the
light that was reflected back along the photonic crystal slab relative
to the energy density of the incident pulse.
This is illustrated in Fig.\ \ref{fdtd}(b), where the energy density of the 
reflected pulse is shown as a function of the 
dielectric constant of the boundary material. 
It was calculated using the two-dimensional FDTD method. 
We can see that with increasing dielectric constant of the 
boundary material more light is reflected along the photonic crystal slab
(indicating that for a totally reflecting boundary material such as metal the
effect would be optimal). This and other properties of
the crystal and its boundaries can be used to optimize the reflection.
We have also studied a slab with a refractive index that is 
the average of that of the photonic crystal slab 
($n=\sqrt{13\cdot 0.2+1\cdot 0.8}$).
The considered boundary materials were dielectric with $\varepsilon_b=13$ 
and metal (Au) with $n=0.053+7.249i$. In these cases only a small fraction
of the light is reflected back along the waveguide. This shows that, for 
this geometry and choice of materials, the effect of the photonic crystal 
cannot be reduced to a homogeneous material with a corresponding average 
refractive index.

Our studies show that in some cases (the 2D periodic structure 
considered here) the guiding of light by patterning the surrounding 
material of a photonic crystal slab seems to reduce to the use of a 
conventional ridge waveguide. However, in other cases (the 1D example), 
the effect of the photonic crystal clearly cannot be reduced to a 
homogeneous material with the average refractive index. The situation 
somewhat resembles the case of microstructured fibers: there, in some 
cases the guiding can be explained by a conventional step index fiber 
structure with an effective refractive index, whereas in other cases the 
guiding is a genuine photonic band gap effect. In both cases, the fibers 
have turned out to provide interesting applications.

The advantages of the guiding scheme proposed here depend on the type of 
guiding it actually provides. In all cases (even when the guiding is 
effectively equivalent to guiding by a homogeneous material ridge 
waveguide), the advantage over using a ridge waveguide is the 
integrability with photonic crystal components, e.g. active elements 
(one may even imagine realizing microcavities by the effect) and 
additional degrees of freedom in designing the mode properties. The 
advantage over defect waveguides may be better guiding in some cases (as 
demonstrated here), easier fabrication or better integrability. In 
addition to these advantages, when the guiding is a genuine photonic 
band gap effect, one should be able to realize, for suitable crystal 
symmetries, sharp bends that are not possible with ridge waveguides. For 
the hexagonal lattice considered here, it was unfortunately not possible 
to demonstrate this since the frequencies of the band gaps as well as 
the frequency of the guided mode differ considerably in the K- and 
M-directions.

In summary, we have studied the effect of boundary materials and 
finite height of one- and two-dimensional photonic crystals to 
the band gap of the photonic crystal. We used a pseudo-spectral 
method and a finite difference time domain method.
The band gap shows dependency on the boundary material. 
The strong effect of the boundary material on the band structures can be 
used for novel type of guiding of light in two-dimensional photonic 
crystals: the channel would be realized by covering a part 
(having the shape of the channel) in otherwise uncovered slab. 
We tested the idea by FDTD simulations of pulse propagation in 
a two-dimensional photonic crystal, and the proposed guiding by 
boundary material provided twice as effective transmission as the 
traditional guiding by a defect, for the (unoptimized) parameters used. 
Whether or not the guiding can be understood by replacing the 
photonic crystal by a homogenous material with the average refractive 
index was shown to depend on the choice of geometry and materials.

\begin{ack}
We thank the Academy of Finland for support (Project Nos.
48845, 53903, 205454). Jerome Moloney and Aramais Zakharian 
(Arizona Center for Mathematical Sciences, University of Arizona) 
are acknowledged for providing the original 3D FDTD code and 
Juuso Olkkonen (VTT Electronics, Finland and Optical Sciences Center, 
University of Arizona) for development and support of the 3D FDTD code.
\end{ack}

\clearpage

\clearpage

\clearpage

\clearpage

\clearpage

\clearpage

\end{document}